\RequirePackage{amsmath}
\documentclass[runningheads]{llncs}

\usepackage[utf8]{inputenc}
\usepackage[T1]{fontenc}
\usepackage[english]{babel}

\usepackage{graphicx}
\usepackage{amssymb,amsfonts}
\usepackage[svgnames,table]{xcolor}
\usepackage{mathtools}
\usepackage{bm}
\usepackage{subcaption}
\usepackage{algorithm}
\usepackage{algpseudocode}
\usepackage[hidelinks]{hyperref}
\usepackage[shortlabels]{enumitem}
\usepackage{listings}
\usepackage{siunitx}
\usepackage{comment}
\usepackage{cite}
\usepackage{booktabs}       
\usepackage{amsmath}
\usepackage{csquotes}

\usepackage{mathtools}
\mathtoolsset{showonlyrefs}

\newcommand{\C}{\mathbb{C}}

\newcommand{\N}{\mathbb{N}}

\newcommand{\BI}{\mathbf{I}}
\newcommand{\BA}{\mathbf{A}}

\newcommand{\Gaussian}{\mathcal{N}}
\newcommand{\Poisson}{\text{Poisson}}

\newcommand{\F}{\mathcal{F}}

\DeclareMathOperator*{\argmin}{arg\,min}

\newcommand{\Bx}{\mathbf{x}}
\newcommand{\By}{\mathbf{y}}
\newcommand{\Bz}{\mathbf{z}}
\newcommand{\diag}{\text{ diag}}

\newcommand{\best}[1]{\cellcolor{black!12}#1}
\newcommand{\iu}{{i\mkern1mu}}

\begin{document}

\title{Plug-and-Play Half-Quadratic Splitting for Ptychography}
%
%
\author{Alexander Denker, Johannes Hertrich, Zeljko Kereta, Silvia Cipiccia, \\ Ecem Erin, Simon Arridge }
\authorrunning{Denker et al.}

\institute{University College London}
%
%
\maketitle              
\begin{abstract}
Ptychography is a coherent diffraction imaging method that uses phase retrieval techniques to reconstruct complex-valued images. It achieves this by sequentially illuminating overlapping regions of a sample with a coherent beam and recording the diffraction pattern. Although this addresses traditional imaging system challenges, it is computationally intensive and highly sensitive to noise, especially with reduced illumination overlap. Data-driven regularisation techniques have been applied in phase retrieval to improve reconstruction quality. In particular, plug-and-play (PnP) offers flexibility by integrating data-driven denoisers as implicit priors. In this work, we propose a half-quadratic splitting framework for using PnP and other data-driven priors for ptychography. We evaluate our method both on natural images and real test objects to validate its effectiveness for ptychographic image reconstruction. 
\end{abstract}

\section{Introduction}
Phase retrieval aims to recover a complex-valued signal $\Bx \in \C^n$ from intensity-only measurements 
\begin{align}
    \By = |\BA \Bx|,
\end{align}
for a linear forward operator $\BA$. The reconstruction of $\Bx$ from $\By$ is a non-linear and ill-posed inverse problem, due to the lack of phase information. 

Phase retrieval is ubiquitous in imaging applications, e.g., X-ray crystallography \cite{sayre1952some} or holography \cite{zhang20173d}, see also the review \cite{dong2023phase}.
Moreover, phase retrieval is a crucial component in many computational microscopy pipelines. Standard microscopy techniques use a lens between the object and the detector to form an image. 
Optical imperfections cause aberrations that limit the resolution of the imaging system. This is particularly severe in the X-ray regime where manufacturing  high-quality optics is both challenging and extremely costly. Coherent diffraction imaging (CDI) offers an alternative by replacing the lens with a computational algorithm, thereby eliminating optical aberrations \cite{candes2015phase}. CDI makes use of diffraction patterns to reconstruct the object of interest. The far-field diffraction patterns can be approximated by the Fourier transform.

\begin{figure}[t]
    \centering
    \includegraphics[width=1.0\linewidth]{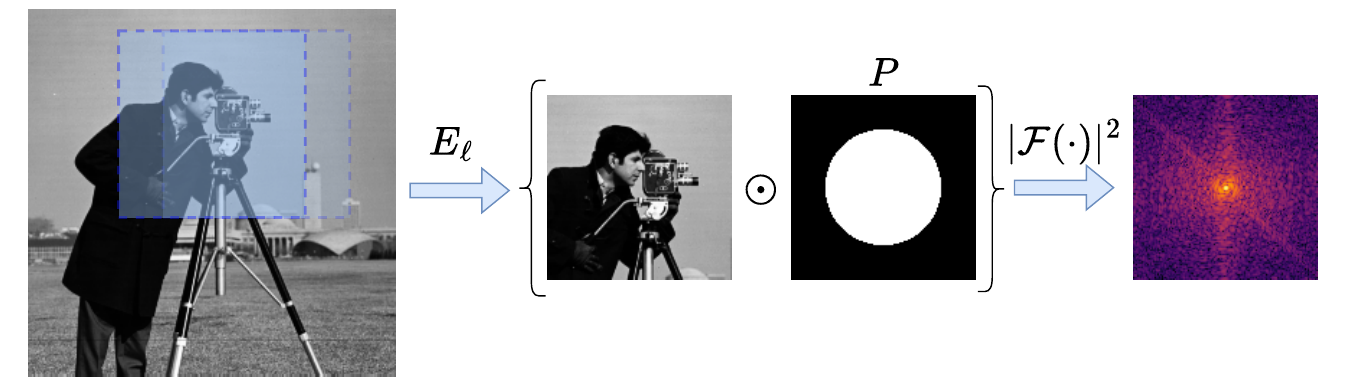}
    \caption{The measurement model for ptychography. We first extract a patch from the image. We take the element-wise product of the extracted patch with the probe. Finally, we obtain the magnitude of the Fourier transform. This process is repeated for all probe positions in the image. The blue squares in the left image show two consecutive overlapping regions.} 
    \label{fig:measurement_setup}
    \vspace{-0.4cm}
\end{figure}

Ptychography is a form of CDI that enables imaging of extended objects by shifting coherent illumination, smaller than the object, across the sample at overlapping positions, see Figure~\ref{fig:measurement_setup} for a visualisation of the measurement procedure. 
For each position, a coherent beam illuminates a small region of the sample, and the intensity of the diffraction pattern from the illuminated region is recorded \cite{Rodenburg_Maiden_2019}. 
The overlap between illuminations introduces redundancy, which mitigates the ill-posed nature of the problem and ensures a stable reconstruction. 
However, increased overlap also increases the number of scanning positions and consequently, the total scanning time. Thus, minimising overlap is desirable. 

Ptychographic reconstruction methods mostly use iterative phase retrieval algorithms \cite{Enders2016,Maiden2009AnIP,Sicairos2008,Wakonig2020}. 
The performance of these algorithms often deteriorates in the presence of high noise and reduced illumination overlap. 
To address these challenges and stabilise the reconstruction process, regularisation techniques can be employed \cite{benning2018modern}. 
Regularisation introduces prior knowledge to improve reconstruction stability and guide it towards a more plausible and stable solution. This can take the form of sparsity-enforcing priors, smoothness constraints, or more sophisticated models that reflect expected properties of the object of interest. 
In particular, data-driven priors like (weakly) convex ridge regularisers \cite{goujon2023neural,goujon2024learning} and patch-priors \cite{altekruger2023patchnr} have seen increased attention, see \cite{habring2024neural} for a recent overview.

A particularly promising approach to leverage data-driven priors is the use of plug-and-play (PnP) algorithms \cite{venkatakrishnan2013plug}. The PnP framework provides a flexible way to incorporate image denoising methods as implicit priors without explicitly defining a regularization term. Building on the success of modern denoisers in phase retrieval \cite{sun2019regularized,wang2020deep,metzler2018prdeep,shastri2024fastrobustphaseretrieval}, we explore the application of PnP to ptychography.


\paragraph{Plug-and-Play} The core concept of PnP algorithms is to implicitly define the regularisation term using a denoiser. PnP algorithms build on classical splitting methods, which separate the data fidelity and regularisation components of the optimisation problem. In these methods, the regularisation term often manifests as a proximal operator, typically corresponding to a Gaussian denoising task.
Hence, the main idea of PnP \cite{venkatakrishnan2013plug} is to replace the proximal operator in variational reconstruction algorithms by a more general denoiser. 
While the authors of \cite{chan2016plug,venkatakrishnan2013plug} used classical denoisers, modern approaches predominantly use neural network-based denoisers  \cite{buzzard2018plug,meinhardt2017learning,zhang2021plug,zhang2017learning}.
Performance of PnP algorithms can be further improved by using dataset-specific denoisers derived from diffusion or flow-matching generative models \cite{graikos2022diffusion,martin2024pnp}. However, this approach reduces the generality of the method.
Regularisation by denoising (RED) \cite{romano2017little} is a concept related to PnP, which does not replace a proximal operator. Instead, it employs gradient descent with a Laplacian-type, differentiable regulariser. 
While PnP methods work well in practice, they usually do not converge, see e.g.~\cite{sommerhoff2019energy} for numerical examples. 
As a remedy, most implementations use only finitely many steps of the underlying reconstruction algorithm. Convergence can be achieved by restricting either the Lipschitz constant or the architecture of the denoiser \cite{hertrich2021convolutional,hurault2022gradient,hurault2022proximal,pesquet2021learning,ryu2019plug}. However, the resulting convergent PnP algorithms typically exhibit slightly worse performance.

\paragraph{Plug-and-Play for Phase Retrieval}
Most existing PnP algorithms were analysed for linear inverse problems. 
However, some algorithms, including PnP with half-quadratic splitting (HQS) and forward-backward splitting (FBS), can also be applied for non-linear inverse problems provided data-fidelity steps can be solved. 
In particular, \cite{sun2019regularized} propose PnP-FISTA for Fourier ptychography and \cite{metzler2018prdeep} propose a PnP-FASTA for phase retrieval. 
Studies show that PnP-HQS usually works better than PnP-FBS and is less sensitive to the step-size parameter \cite{zhang2021plug}.
Following this line of research, we propose using PnP-HQS for phase retrieval and ptychography. In Section~\ref{sec:hqs-pt}, we show that the corresponding data-fidelity steps can be explicitly solved. 
Further, most PnP algorithms for phase retrieval only deal with real-valued images, see \cite{sun2019regularized,metzler2018prdeep,wang2020deep,shastri2024fastrobustphaseretrieval}, and exploit pre-trained denoisers for natural grey-scale images. However, objects of interest in ptychography are inherently complex-valued with the amplitude and phase encoding important physical information. 
Thus, we require denoisers for complex-valued images. 
\vspace{-0.4cm}

\subsection*{Mathematical Model of the Forward Operator}
The measurement process in ptychography, as visualised in Fig.~\ref{fig:measurement_setup}, can be formally described by defining a window operator (coded illumination) $E_\ell$, for $\ell\in\{1,...,L\}$, which extracts a $N\times N$ window around a probe position $p_\ell$.  
Measurements $\By_\ell$ are generated by first applying the window extraction operator $E_\ell$ followed by an element-wise multiplication with $P$, representing the illumination function or a mask. Finally, the wave is transported to the far-field via the Fourier transform and its magnitude is recorded.
This process can be summarised by
\begin{equation}
    \By_\ell^2 = | \F(A_{\ell} \Bx) |^2 \quad \ell=1,\dots,L, \label{eq:ptychography_forward}
\end{equation}
where $A_\ell=P\circ E_\ell$ is the concatenation of the probe with the window extraction operator, and $\F$ is the 2D Fourier transform.
For notational simplicity, in the following we identify all the images, such as $\Bx$ or $\By$, through their vectorisations. 
This allows representing $A_\ell$ as diagonal matrices (since they encode element-wise operations), with adjoints denoted $A_\ell^*$. 
Thus, $A_\ell^*A_\ell$ are diagonal matrices. 

\paragraph{Noise Model and Data-Fidelity Term}
We use a shot noise model as in \cite{metzler2018prdeep}
\begin{align}\label{eq:noise_model}
    \By_\ell^2 := | \F(A_{\ell} \Bx) |^2 + \alpha  \diag(| \F(A_{\ell} \Bx) |^2) \eta, \text{ for } \eta \sim \Gaussian(0, \BI).
\end{align}
This approximates a Poisson noise model with $\By_\ell^2 / \alpha^2 \sim \Poisson(| \F(A_{\ell} \Bx) |^2 / \alpha^2)$, where $\alpha$ controls the signal-to-noise ratio.
Even though we do not deal with Gaussian noise, we follow \cite{yeh2015experimental} and use $\frac12\|\By-G(\Bz)\|^2$ as the data-fidelity term, where $G$ and $\By$ are the forward operator and measurement vector obtained by stacking the forward operators $|(\F\circ A_\ell)(\cdot)|$ and amplitudes $|\F(A_\ell\Bx)|$ of each probe. This simplification is widely used in ptychography-related applications~\cite{metzler2018prdeep,yeh2015experimental}.

\section{Half-Quadratic Splitting}\label{sec:hqs-pt}
In this section we derive a half-quadratic splitting (HQS) algorithm \cite{geman1995nonlinear} for ptychography. To this end, we consider the variational regularisation problem
\begin{equation}\label{eq:variational_problem}
\argmin_{\Bx}\frac{1}{2}\|\By-G(\Bx)\|^2+\lambda R(\Bx),
\end{equation}
where $R$ is a convex regularisation term, $\lambda>0$ a regularisation parameter, and $G$ is either the forward operator \eqref{eq:ptychography_forward}, in case of ptychography, or $G(\Bx)=|\F(\Bx)|$,  in case of phase retrieval. 
We first focus on phase retrieval, where we show that data-consistency steps can be solved analytically. 
We then turn our focus to ptychography, where we introduce multiple auxiliary variables and reduce the resulting subproblems to the phase retrieval case.

\subsection{Half-Quadratic Splitting for Fourier Phase Retrieval}
\label{sec:HQS_PR}
For $G(\Bx) = |\F(\Bx)|$, the variational problem \eqref{eq:variational_problem} can be reformulated as
$$
\argmin_{\Bx,\Bz}\frac{1}{2}\|\By-|\F(\Bz)|\|^2+\lambda R(\Bx),\quad \text{s.t.}\quad\Bx=\Bz.
$$
Relaxing the constraint $\Bx=\Bz$ by a quadratic penalisation, we obtain the problem
$$
\argmin_{\Bx,\Bz}\mathcal L_\mu(\Bx,\Bz)\coloneqq\frac{1}{2}\|\By-|\F(\Bz)|\|^2+\frac{\mu}{2}\|\Bz-\Bx\|^2+\lambda R(\Bx).
$$
The original problem is recovered by letting $\mu\to\infty$. In the following, for its solution we consider the alternating minimisation method
\begin{align}
    \Bz_k &= \argmin_\Bz \| \By - |\F(\Bz)| \|^2 + \mu \| \Bz - \Bx_{k-1} \|^2, \label{eq:hsq_pr_1} \\ 
    \Bx_k &= \argmin_\Bx \frac{\mu}{2\lambda} \| \Bx - \Bz_k \|^2 + R(\Bx)=\mathrm{prox}_{\frac{\lambda}{\mu} R}(\Bz_k). \label{eq:hsq_pr_2}
\end{align}
Whereas for linear inverse problems, subproblem \eqref{eq:hsq_pr_1} has an efficient closed form solution, for non-linear inverse problems this can be challenging to solve. 
However, in the case of Fourier phase retrieval, we can exploit the structure of the forward operator to derive an explicit solution. Substituting $\Bz=\F^{-1}(\hat \Bz)$, subproblem~\eqref{eq:hsq_pr_1} is equivalent to
\begin{align}
\Bz_k=\F^{-1}(\hat \Bz_k),\quad \hat \Bz_k&=\argmin_{\hat \Bz}\|\By-|\F(\F^{-1}(\hat \Bz))|\|^2+\mu\|\F^{-1}(\hat \Bz)-\Bx_{k-1}\|^2\\
&=\argmin_{\hat \Bz}\|\By-|\hat \Bz|\|^2+\frac{\mu}{n}\|\hat \Bz-\F(\Bx_{k-1})\|^2,\label{eq:hsq_pr2_sub}
\end{align}
where the equality comes from the fact that $\frac{1}{\sqrt{n}}\F$ is an isometry, with $n$ representing the number of components (pixels) of $\Bx$.
In the following, we denote $\hat \Bx_{k-1}=\F(\Bx_{k-1})$. The objective in \eqref{eq:hsq_pr2_sub} is fully separable with respect to the argument.
In particular, the minimiser $\hat\Bz_k=(\hat z_{k,i})_{i=1}^n$ is given by
\begin{align}
    \hat z_{k,i}=\argmin_{\hat z_i} \,(y_i-|\hat z_i|)^2+\frac{\mu}{n}|\hat z_i-\hat x_{k-1,i}|^2.
\end{align}
Since the first term is independent from the phase of $\hat z_i$ and the second term is minimised for $\mathrm{phase}(\hat z_i)=\mathrm{phase}(\hat x_{k-1,i})$ for any fixed amplitude $|\hat z_i|$, we have that $\mathrm{phase}(\hat z_{k,i})=\mathrm{phase}(\hat x_{k-1,i})$. 
It remains to minimise with respect to the amplitude of $\hat z_{k,i}$. Inserting the solution with respect to the phase into the above minimisation problem, the amplitude $|\hat z_{k,i}|$ is given by
\begin{align*}
|\hat z_{k,i}|=\argmin_{r} (y_i-r)^2+\frac{\mu}{n}(r-|\hat x_{k-1,i}|)^2=c y_i+(1-c)|\hat x_{k-1,i}|,
\end{align*}
where $c=\frac{n}{n+\mu}$. In summary, the solution  is given by  
\begin{equation}\label{eq:sol_first_step}
\Bz_k=\F^{-1}(\hat \Bz_k),\text{ for }\hat \Bz_k=(c \By+(1-c)|\hat \Bx_{k-1}|)\exp(\iu\, \mathrm{phase}(\hat \Bx_{k-1})),
\end{equation}
where $c=\frac{n}{n+\mu\sigma^2}$, $\hat \Bx_{k-1}=\F(\Bx_{k-1})$ and all operations are applied pointwise.

\subsection{Multiple Auxiliary Variables in HQS for Ptychography}

For ptychography, we reformulate the variational problem \eqref{eq:variational_problem} as
\begin{align*}
    \argmin_{\Bx,\Bz}\sum_{\ell=1}^L\frac{1}{2}\|\By_\ell-|\F(\Bz_\ell)|\|^2+\lambda R(\Bx),\quad \text{s.t.}\quad A_\ell\Bx=\Bz_\ell.
\end{align*}
Relaxing the constraints, we aim to minimise
\begin{align*}
\mathcal J_\mu(\Bx,\Bz)\coloneqq\sum_{\ell=1}^L\frac{1}{2}\|\By_\ell -|\F(\Bz_l)|\|^2+\sum_{\ell=1}^L\frac{\mu}{2}\|\Bz_\ell-A_\ell\Bx\|^2+\lambda R(\Bx),
\end{align*}
where the original problem formulation is recovered in the limit $\mu\to\infty$.
Applying an alternating minimisation scheme, this leads to the steps
\begin{align}
    \Bz_{k,\ell} &= \argmin_\Bz \| \By_\ell - |\F(\Bz_\ell)| \|^2 + \mu \| \Bz_\ell - A_\ell\Bx_{k-1} \|^2,\quad \ell=1,...,L \label{eq:hsq_pt_1} \\ 
    \Bx_k &= \argmin_\Bx \frac{\mu}{2 \lambda}\sum_{\ell=1}^L\| A_\ell\Bx - \Bz_{k,\ell} \|^2 + R(\Bx). \label{eq:hsq_pt_2}
\end{align}
The first step \eqref{eq:hsq_pt_1}, is the same as the first step \eqref{eq:hsq_pr_1} for phase retrieval and admits an explicit solution given by \eqref{eq:sol_first_step} for each $\ell$.
To solve the second step \eqref{eq:hsq_pt_2}, we first rewrite the sum of the squared norms as
\begin{align}
\sum_{\ell=1}^L\| A_\ell\Bx - \Bz_{k,\ell} \|^2&\propto\Bx^* \left(\sum_{\ell=1}^L |A_\ell|^2\right)\Bx-2\left(\sum_{\ell=1}^L\Bz_{k,\ell}^* A_\ell\right)\Bx,\\
&\propto \left\|\left(\sum_{\ell=1}^L |A_\ell|^2\right)^{\frac12}\Bx - \left(\sum_{\ell=1}^L |A_\ell|^2\right)^{-\frac12} \left(\sum_{\ell=1}^LA_\ell^*\Bz_{k,\ell}\right)\right\|^2, \label{eq:spatially_varying_problem}
\end{align}
where $\propto$ indicates equality up to additive constants independent of $\Bx$, and we use the diagonality of $A_\ell$ to establish $A_\ell^*A_\ell=|A_\ell|^2$.
Here, $\sum_{\ell=1}^L |A_\ell|^2$ corresponds to the accumulated squared intensity of a probe hitting a pixel, and $\left(\sum_{\ell=1}^L |A_\ell|^2\right)^{-\frac12} \left(\sum_{\ell=1}^LA_\ell^*\Bz_{k,\ell}\right)$ is the average of vectors $\Bz_{k,\ell}$ weighted with respect to probe matrices $A_\ell$.
Inserting \eqref{eq:spatially_varying_problem} and denoting the diagonal matrix $D=\left(\sum_{\ell=1}^L |A_\ell|^2\right)^{\frac12}$, and $\tilde \Bz_k=\frac{\sum_{\ell=1}^LA_\ell^*\Bz_{k,\ell}}{\sum_{\ell=1}^L |A_\ell|^2}$, we can reformulate \eqref{eq:hsq_pt_2} as
\begin{align}
    \label{eq:spatially_varying_denoiser}
    \Bx_k = \argmin_\Bx \frac{\mu}{2\lambda }\| D\Bx - D\tilde \Bz_k \|^2 + R(\Bx)=D^{-1}{\rm prox}_{\frac{\lambda}{\mu} R\circ D^{-1}}(D\tilde\Bz_k).
\end{align}
This corresponds to a denoising problem with spatially varying noise level, which has been used in PnP algorithms \cite{pendu2023preconditioned}, albeit in a different context.

\begin{algorithm}[t]
    \begin{algorithmic}
        \State \textbf{Input:} Measurements $(\By_\ell)_{\ell=1}^L$, denoising parameters $(\tau_k)_{k\in\N}$, regularisation strength~$\lambda$, initialization $\Bx_0$
        \For{$k=1,2,...$}
        \State $\mu_k = \lambda / \tau_k^2$
            \State $c_k=\frac{n}{n+\mu_k}$
            \For{$\ell=1,...,L$}
                \State $\hat \Bx_{k-1,\ell}=\F(A_\ell\Bx_{k-1})$
                \State $\Bz_{k,\ell}=\F^{-1}(\hat \Bz_{k,\ell})$ with $\hat \Bz_{k,\ell}=(c_k \By_\ell+(1-c_k)|\hat \Bx_{k-1,\ell}|)\exp({\rm i}\, \mathrm{phase}(\hat \Bx_{k-1,\ell}))$
            \EndFor
            \State $\tilde \Bz_k=\frac{\sum_{\ell=1}^LA_\ell^*\Bz_{k,\ell}}{\sum_{\ell=1}^L |A_\ell|^2}$
            \State $\Bx_k = D^{-1}{\rm prox}_{\tau_k R\circ D^{-1}}(D\tilde\Bz_k) \quad $  
        \EndFor

    \end{algorithmic}
    \caption{Half-quadratic splitting algorithm for Ptychography}
    \label{alg:hqs-pt}
\end{algorithm}

\begin{remark}\label{rem:no_mask}
    When the entries of the probe and the probe positions are sufficiently regular we can assume $D^2\approx \gamma I$ for some $\gamma>0$ such that
$$
\sum_{\ell=1}^L\| A_\ell\Bx - \Bz_{k,\ell} \|^2\approx \gamma\left\|\Bx - \tilde \Bz_k\right\|^2 \quad\text{with}\quad \tilde \Bz_k=\frac{\sum_{\ell=1}^LA_\ell^*\Bz_{k,\ell}}{\sum_{\ell=1}^L |A_\ell|^2},
$$
where division is understood element-wise. Thus, problem~\eqref{eq:hsq_pt_2} can be written as
\begin{equation}\label{eq:sol_second_step}
\Bx_k = \argmin_\Bx \frac{\mu\gamma}{2\lambda }\| \Bx - \tilde \Bz_{k} \|^2 + R(\Bx)=\mathrm{prox}_{\frac{1}{\mu\gamma}R}(\tilde \Bz_k).
\end{equation}
In particular, when $R(\Bx) = \iota_{[0, \infty)^n}(\Bx)$ is the indicator function of the non-negative orthant, we recover the commonly used simultaneous PIE algorithm \cite{konijnenberg2016combining} for ptychography and the error reduction algorithm \cite{fienup1982phase} for phase retrieval.
\end{remark}

In practice, we choose an increasing sequence $(\mu_k)_k$ and for each $k$ perform only one step of HQS. To determine $\mu_k$ we follow a similar principle to that in \cite{zhang2021plug}. For this, note that if we replace the proximal mapping with a learned denoiser, the denoising strength is given by $\tau_k = \sqrt{\lambda/\mu_k}$. As the denoisers are trained for a specific range of noise levels, it is easier to choose $\tau_k$ due to a clear interpretation. Choosing $\tau_k$ defines a sequence $\mu_k = \lambda / \tau_k^2$ with $\lambda$ as the remaining hyper-parameter. We choose $\Tilde{\lambda}$ by $\lambda = \Tilde{\lambda} \hat{\sigma}^2$, where $\hat{\sigma}^2$ is an estimate of the noise level. This results in the algorithm summarised in Algorithm~\ref{alg:hqs-pt}.

\subsection{Complex Denoiser}
\label{sec:complexdenoiser}
To achieve a complex-valued reconstruction, we have several ways for applying the proximal mapping, depending on the choice of the regulariser. Decomposing into real and imaginary parts, or into magnitude and phase, we can apply a denoiser pre-trained on real-valued greyscale images. For instance, for a regulariser of the form $R(\Bz) = R_1(\Re(\Bz)) + R_2(\Im(\Bz))$ we can write the proximal mapping as 
\begin{align}
 {\rm prox}_R(\Bz) = {\rm prox}_{R_1}(\Re(\Bz)) + \iu{\rm prox}_{R_2}(\Im(\Bz)),
\end{align}
due to the separability of the proximal mapping. This motivates to apply a pre-trained denoiser to the real and imaginary parts independently. To ensure non-negativity of the real and imaginary parts, before denoising we add a constant equal to the maximum amplitude. This pre-processing step is required as most pre-trained denoisers are designed to work only for non-negative images.



\section{Experiments}
We compare PnP approaches against two commonly used classical algorithms: Sequential PIE (SeqPie) and Simultaneous PIE (SimPie) \cite{konijnenberg2016combining}. All PnP algorithms are initialised with 100 iteration of SimPIE. Moreover, we compare different choices of denoisers for PnP algorithms. First, we use two regularisers for which we solve the spatially varying proximal mapping as in Eqn.~\eqref{eq:spatially_varying_denoiser}, applied independently to the real and imaginary components of the image. In particular, we use Total Variation (TV) \cite{rudin1992nonlinear} and the weakly convex ridge regulariser (WCRR) \cite{goujon2024learning}. 
Finally, we use a pre-trained DRUNet \cite{zhang2021plug}\footnote{Implementation from the \texttt{deepinv} package \url{https://deepinv.github.io}} and use the equivariant evaluation from \cite{terris2024equivariant}. For the pre-trained DRUNet, we use the approximation from Remark~\ref{rem:no_mask}. 
We compute the PSNR on the amplitude and on the phase. For the phase, we correct for a global phase shift and calculate the phase PSNR as 
\begin{align}
    \text{PSNR}_\phi(\phi_\text{reco}, \phi_\text{gt}) = 10 \log_{10}\left(\frac{(2 \pi)^2}{\text{MSE}_\phi(\phi_\text{reco}, \phi_\text{gt})}\right),
\end{align}
where $\text{MSE}_\phi$ is the mean squared error on the sphere defined as
\begin{align*}
    \text{MSE}_\phi(\phi_\text{reco}, \phi_\text{gt}) = \frac{1}{N} \sum_{i=1}^N ((\phi_{\text{reco},i} - \phi_{\text{gt},i} + \pi)\bmod{2\pi} - \pi)^2,
\end{align*}
and $\phi_\text{reco}$ and $\phi_\text{gt}$ are the phases of the reconstruction and ground truth image. Because measurements are sparse at the border, it cannot be reconstructed. Therefore, we exclude a 20-pixel border and focus on the center of the image when calculating the quality metrics.

\begin{figure}[t]
    \centering
    \includegraphics[width=1.0\linewidth]{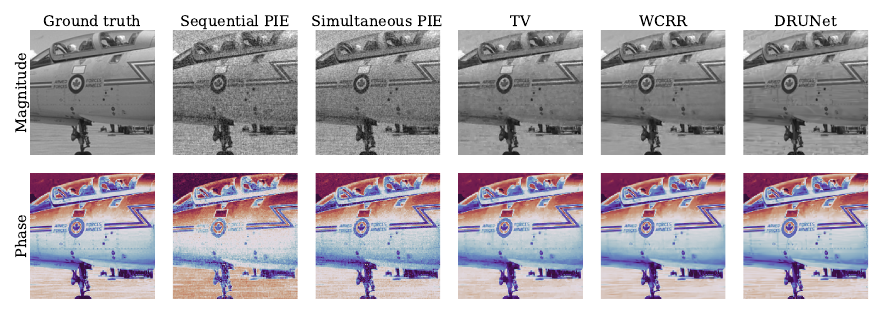}
    \caption{Reconstructions for the $7\times 7$ probe setting with $\alpha=20.0$. We only show the center of the image where PSNR is calculated.}
    \label{fig:example_reco}
    \vspace{-0.6cm}
\end{figure}

\subsection{Natural Images}
We evaluate our choices of denoisers on the BSD500 dataset \cite{MartinFTM01} of natural RGB colour images. We first crop the central $256 \times 256$px region, and create a complex valued image by setting the amplitude to be the mean of the red and green channels. For the phase we scale the blue colour channel to $[-\pi, \pi]$ and add a random global phase shift. We use a binary circular probe with a radius of $40$px, cf.~the experimental setup in Fig.~\ref{fig:measurement_setup}. The probe positions are evenly distributed. For all methods we use $600$ iterations and choose hyperparameters to maximise the amplitude PSNR on a small validation set for $38\%$ overlap and $\alpha=20$. For TV we choose denoising strength in $[6.0, 30.0]$ and $\lambda=0.0001$. For the WCRR we choose  denoising strengths in $[200, 1500]$ and $\lambda=0.01$. Finally, for the DRUNet we use denoising strengths in $[5.0, 30.0]$ and $\lambda=0.0001$. For SimPIE and SeqPIE, we used $2000$ iterations. The same hyperparameters were used for all settings.

\vspace{-0.2cm}
\paragraph{Influence of the Overlap} We evaluate how the overlap percentage, determined by the number of probe positions, affects the performance. We test with overlap percentages from $38\%$ (49 probe locations) to $68\%$ (225 probe locations). With classical algorithms (e.g. PIE) an overlap of $60 - 70\%$ is required to produce good results. For binary probes, overlap is computed as the intersection over union of the support. 
Results are presented in Table~\ref{tab:phase_retrieval_results_probes} and an example reconstruction is shown in Fig.~\ref{fig:example_reco}. For all probe setting we consistently see that HQS with different denoisers outperforms the (unregularised) PIE. In particular, the performance gap is bigger for a smaller overlap. The neural network based regularisers outperform the TV regulariser for all settings. 

\begin{table}[t]
\centering
\caption{PSNR (mean $\pm$ standard deviation) for amplitude and phase for the complex-valued BSD500 images with a noise level $\alpha = 20.0$ for different probe setups with a circular probe. Best results are highlighted. }
\resizebox{\textwidth}{!}{%
\begin{tabular}{lcccccccccccc}
       \toprule Probes && \multicolumn{2}{c}{$7\times 7$ ($38\%$ overlap)}   && \multicolumn{2}{c}{$9\times 9$ ($48\%$ overlap)} && \multicolumn{2}{c}{$11 \times 11$ ($56\%$ overlap)} && \multicolumn{2}{c}{$15 \times 15$ ($68\%$ overlap)} \\
      \cmidrule{3-4} \cmidrule{6-7} \cmidrule{9-10} \cmidrule{12-13}
 && PSNR$_a$   & PSNR$_\phi$ && PSNR$_a$   & PSNR$_\phi$  && PSNR$_a$   & PSNR$_\phi$ && PSNR$_a$   & PSNR$_\phi$  \\ \midrule
SeqPIE   && $21.36${\scriptsize $\pm 1.30$}  & $23.03${\scriptsize $\pm 5.68$}  && $21.93${\scriptsize $\pm 1.22$} & $23.85${\scriptsize $\pm 5.53$}  && $21.99${\scriptsize $\pm 1.18$}  & $23.50${\scriptsize $\pm 5.70$} && $21.91${\scriptsize $\pm 1.17$} & $23.78${\scriptsize $\pm 5.64$}  \\
SimPIE && $23.29${\scriptsize $\pm 1.25$}  & $25.35${\scriptsize $\pm 5.95$}  && $26.25${\scriptsize $\pm 0.99$} & $28.08${\scriptsize $\pm 6.50$}  && $28.39${\scriptsize $\pm 0.92$}  & $30.10${\scriptsize $\pm 6.86$} && $31.40${\scriptsize $\pm 0.99$} & $33.27${\scriptsize $\pm 7.16$}  \\
TV && $26.13${\scriptsize $\pm 2.85$}  & $28.72${\scriptsize $\pm 6.11$}  && $28.90${\scriptsize $\pm 2.03$} & $31.33${\scriptsize $\pm 6.28$}  && $30.56${\scriptsize $\pm 1.63$}  & $32.81${\scriptsize $\pm 6.70$} && $32.70${\scriptsize $\pm 1.63$} & $35.19${\scriptsize $\pm 6.88$}  \\
DRUNet  && \best{$27.73${\scriptsize $\pm 2.35$}}  & $29.76${\scriptsize $\pm 6.65$}  && \best{$30.24${\scriptsize $\pm 2.07$}} & $32.11${\scriptsize $\pm 6.63$}  && $31.07${\scriptsize $\pm 2.63$}  & $32.85${\scriptsize $\pm 6.87$} && $32.00${\scriptsize $\pm 2.10$} & $34.30${\scriptsize $\pm 5.70$}  \\
WCRR  && $27.03${\scriptsize $\pm 2.90$}  & \best{$30.01${\scriptsize $\pm 5.93$}}  && $29.56${\scriptsize $\pm 2.10$} & \best{$32.38${\scriptsize $\pm 6.46$}}  && \best{$31.13${\scriptsize $\pm 1.84$}}  & \best{$33.86${\scriptsize $\pm 6.20$}} && \best{$33.12${\scriptsize $\pm 1.84$}} & \best{$35.81${\scriptsize $\pm 6.99$}}  \\

 \bottomrule          
\end{tabular}}
\label{tab:phase_retrieval_results_probes}
\vspace{-0.3cm}
\end{table}

\vspace{-0.3cm}
\paragraph{Influence of Noise} For a fixed probe setting, we evaluate the performance of different algorithms under varying noise levels. Here, we use the $7 \times 7$ probe setting, which results in a overlap of $38 \%$. We test noise levels $\alpha=10,20,30,40$. The results are in Table~\ref{tab:phase_retrieval_results_probes}. Similar to previous results, PnP approaches outperform SimPIE. Among the data-driven denoisers, the WCRR has a slightly better performance compared to the DRUNet for higher noise levels. Whereas the DRUNet and TV result in a similar performance at this level. For the unregularised SimPIE we observe a bigger drop in performance with increasing noise level.

\begin{table}[t]
\centering
\caption{PSNR (mean $\pm$ standard deviation) for amplitude and phase for the complex-valued BSD500 images for $7 \times 7$ circular probes ($38\%$ overlap) for different noise levels. Best results are highlighted.}
\resizebox{\textwidth}{!}{%
\begin{tabular}{lcccccccccccc}
       \toprule Noise level && \multicolumn{2}{c}{$\alpha = 10.0$}   && \multicolumn{2}{c}{$\alpha = 20.0$} && \multicolumn{2}{c}{$\alpha = 30.0$} && \multicolumn{2}{c}{$\alpha = 40.0$} \\
      \cmidrule{3-4} \cmidrule{6-7} \cmidrule{9-10} \cmidrule{12-13}
 && PSNR$_a$   & PSNR$_\phi$ && PSNR$_a$   & PSNR$_\phi$  && PSNR$_a$   & PSNR$_\phi$ && PSNR$_a$   & PSNR$_\phi$  \\ \midrule
SimPIE && $28.79${\scriptsize $\pm 0.79$}  & $30.89${\scriptsize $\pm 6.45$}  && $23.29${\scriptsize $\pm 1.25$} & $25.35${\scriptsize $\pm 5.95$}  && $20.39${\scriptsize $\pm 1.61$}  & $22.43${\scriptsize $\pm 5.33$} && $18.61${\scriptsize $\pm 1.85$} & $20.76${\scriptsize $\pm 4.87$}  \\
TV && $28.09${\scriptsize $\pm 2.32$}  & $30.16${\scriptsize $\pm 6.21$}  && $26.13${\scriptsize $\pm 2.85$} & $28.72${\scriptsize $\pm 6.11$}  && $23.39${\scriptsize $\pm 3.29$}  & $26.17${\scriptsize $\pm 6.08$} && $21.18${\scriptsize $\pm 3.35$} & $24.23${\scriptsize $\pm 5.67$}  \\
DRUNet  && \best{$32.35${\scriptsize $\pm 2.13$}}  & \best{$34.89${\scriptsize $\pm 5.98$}}  && \best{$27.73${\scriptsize $\pm 2.35$}} & $29.76${\scriptsize $\pm 6.65$}  && $24.01${\scriptsize $\pm 2.43$}  & $25.79${\scriptsize $\pm 6.42$} && $21.47${\scriptsize $\pm 2.57$} & $23.35${\scriptsize $\pm 5.72$}  \\
WCRR && $31.20${\scriptsize $\pm 2.42$}  & $33.92${\scriptsize $\pm 5.87$}  && $27.03${\scriptsize $\pm 2.90$} & \best{$30.01${\scriptsize $\pm 5.93$}}  && \best{$24.28${\scriptsize $\pm 2.98$}}  & \best{$27.38${\scriptsize $\pm 5.47$}} && \best{$22.17${\scriptsize $\pm 3.04$}} & \best{$25.35${\scriptsize $\pm 5.18$}}  \\
 \bottomrule          
\end{tabular}}
\label{tab:phase_retrieval_results_noise}
\vspace{-0.4cm}
\end{table}

\vspace{-0.25cm}
\subsection{Brain Phantom}
\begin{figure}[th]
    \vspace{-0.4cm}
    \centering
    \includegraphics[trim={0 0.3cm 0 0.25cm},clip,width=1.0\linewidth]{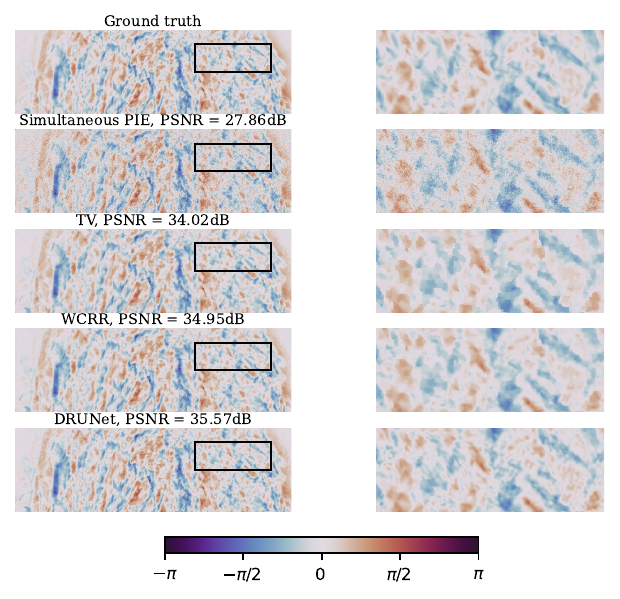}
    \caption{Reconstruction of the phase of the brain phantom. The ground truth is given as the converged ePIE \cite{Maiden2009AnIP} solution from oversampled measured data.}
    \label{fig:brain_phantom}
    \vspace{-0.3cm}
\end{figure}

Finally, we employ the PnP algorithms to a brain phantom with a resolution of $800\times 2200$px. We simulate measurements using a circular probe, noise level $\alpha=40$ and a probe setting of $57\times23$, resulting in an overlap of $36.3\%$ in the x-axis and $44.7\%$ in the y-axis. The phantom has a low constant amplitude, so we only show the phase in Fig.~\ref{fig:brain_phantom}. For this phantom, PnP regularisers again outperform SimPIE. The TV reconstruction exhibits the typical stair-casing effect, whereas both the WCRR and the DRUNet reconstruction appear more realistic, despite being trained on natural images.  

\section{Conclusion and Further Work}
In this work we extended the PnP-HQS algorithm to ptychographic image reconstruction. We show that the data-fidelity step can be solved explicitly, reducing the computational cost. In our initial experiments, the PnP-HQS algorithm was able to produce good reconstructions while using lower probe overlap than classical methods. Thus, the developed framework could be a first step to reducing the acquisition time. 
The next important step in applying this methodology to realistic ptychography systems is to reconstruct both the object and the probe (which is in this work assumed to be known) simultaneously \cite{Maiden2009AnIP}. 


%
%
%
\bibliographystyle{splncs04}
\bibliography{bib}
%




\end{document}